\documentclass[a4paper]{jpconf}
\usepackage{iopams}
\usepackage{lipsum,babel}
\usepackage{graphics}
\usepackage{color}
\usepackage[usenames,dvipsnames]{xcolor}
\usepackage{dcolumn}
\usepackage{bm}
\usepackage[latin1]{inputenc}
\usepackage{textcomp}
\usepackage{footnote}
\usepackage{epstopdf}
\usepackage{url}
\usepackage{amssymb}
\usepackage{balance}
\usepackage{cite}
\usepackage{hyperref}
\usepackage{gensymb}
\def \MgB2 {MgB$_{2}$ }
\begin{document}

\title[3D Modelling and Validation of Commercial CORC Cable]{3D Modelling and Validation of the Optimal Pitch in Commercial CORC Cables}

\author{M. U. Fareed, M. Kapolka, B. C. Robert, M. Clegg, \& H. S. Ruiz}
\address{College of Science and Engineering \& Space Park Leicester, University of Leicester, Leicester LE1 7RH, United Kingdom}
\ead{muf2@le.ac.uk ; mk652@le.ac.uk ; hsrr1@le.ac.uk } 


\begin{abstract}
Conductor on a rounded core (CORC\textsuperscript{\textregistered}) cables with current densities beyond 300 A/mm$^{-2}$ at 4.2 K, and a capacity to retain around 90 $\%$ of critical current after bending to a diameter of 3.5 cm, make them a strong candidate for high field power applications and magnets. In this paper, we present a full 3D-FEM model based upon the so-called H-formulation for commercial CORC\textsuperscript{\textregistered} cables manufactured by Advanced Conductor Technologies LLC. The model presented consists of tapes ranging from 1 up to 3 SuperPower 4mm-width tapes in 1 single layer and at multiple pitch angles. By varying the twist pitch, local electromagnetic characteristics such as the current density distribution along the length and width are visualized. Measurements of macroscopical quantities such as AC-losses are disclosed in comparison with available experimental measurements. We particularly focused on the influence of the twist pitch by comparing the efficiency and performance of multiple cables, critically assessing the optimal twist pitch angle. 
\end{abstract}

%
%
%
%

\section{Introduction}
\label{Sec.1}

The future updating of large-scale superconducting magnets for high-energy applications, such as those involved in the High-Luminosity Large Hadron Collider (HL-LHC) project~\cite{HL_LHC}, and fusion energy projects such as ITER~\cite{ITER}, both require the optimal designing of high current capacity cables for  minimizing the number of Ampere-turns in the manufacturing of superconducting coils. Thus, in the last few years a very promising design of high temperature superconducting cables has started to dominate the landscape for high field applications~\cite{Wang2018SUST} by following the simple concept of winding Coated Conductors On a Round Core, nowadays known as CORC\textsuperscript{\textregistered} cable~ \cite{vanderLaan2009SUST}. These coated conductors are also known as the second generation of high temperature superconducting (2G-HTS) tapes, which when helically wounded on a cylindrical core offer superior advantages over other cabling technologies such as,  a low manufacturing cost, a relatively easy production, and enhanced engineering current density~\cite{Souc2017SUST}. They also offer the possibility to have a full transposition of the tapes without affecting their physical properties along the cable length, it because of the relatively high elasticity module of the 2G-HTS tapes, and the use of flexible formers with a small diameter. It allows the cable structure to bend in any direction without irreversible degradation \cite{vanderLaan2007APL}

The potential of CORC\textsuperscript\textregistered cables for high field magnets have been gradually investigated since their inception, performing viability studies on one to multiple layer cables at up to $20$~T, and at different temperatures~\cite{vanderLaan2016SUST,Mulder2020IEEE}. These studies have demonstrated engineering current densities beyond the $300$~Amm$^{-2}$ at 4.2~K and up to 20 T \cite{Mulder2018IEEE,vanderLaan2019SUST}, proving the concept of CORC\textsuperscript\textregistered cables as a suitable candidate for future high field magnets. Moreover, it has been recently reported that CORC\textsuperscript\textregistered cables with low tape windings on strong elastic cores are indeed suitable for the manufacturing of 40-60 T field magnets, allowing an effective protection against the high operating stresses (>600 MPa) and strains associated with high-field magnet operation~\cite{vanderLaan2021SUST}. Nevertheless, by reducing the tape winding angle, the amount of tape necessary to manufacture a determined length of a conductor can increase substantially and furthermore, it is already known that variations of the angle with which each one of the tapes is wound onto the former could severely change the magnetization measurements of the CORC\textsuperscript\textregistered cable~\cite{Katsutoku2011SUST,Goo2021IEEE}. Thus, this can lead to the possibility to encounter robust increments on the AC-losses of the cable once in operation. Nevertheless, it is known that the magnetization losses of the overall cable can be reduced by optimal twisting (pitch angle) of the multiple tapes at a single layer, which should be first computationally assessed in order to reduce the prototyping and manufacturing costs.

Due to the rising complexity and demand of CORC\textsuperscript\textregistered cables with applications pressing for a higher number of layers and 2G-HTS tapes, a deep understanding and sound computational model for reproducing the electromagnetic characteristics of commercially available CORC\textsuperscript{\textregistered} cables, and in particular its AC-losses, results are of utter importance. Nevertheless, the high aspect ratio of the helically twisted tapes at the CORC\textsuperscript\textregistered cable, makes the devising of finite element models computationally demanding. Although with the continuous improvement of software and hardware, every year this becomes less of a concern. Certain simplifications can still be useful as long as the electromagnetic quantity of interest is not the AC-losses, as the level of accuracy demanded in this case is commonly the highest. This is added to the fact that the change on the power density that leads to the superconducting energy losses cannot be fully captured by 2D models, as it is the motion of the profiles of current density inside the superconducting tapes what causes the losses. In this sense, 2D models of CORC\textsuperscript\textregistered cables~\cite{Li2014IEEE} do not consider the twist pitch, and hence at short twist pitch lengths they can suffer of significant errors. By using symmetry conditions at the full 3D geometry whose approaches can render to computations over a 2D domain, these drawbacks can be addressed. Assuming that the current only flows along the 2D domains following the helicoidal trajectories, a scalar power-law for the resistivity of the superconductor makes the eddy current approach feasible for the solution of the 3D CORC\textsuperscript\textregistered cable but within a 2D formulation for the electromagnetic variables~\cite{Stenvall2013SUST}. Similarly, by neglecting any significant hysteretic process across the thickness of the 2G-HTS tapes, i.e., by assuming infinitely thin superconducting tapes where all relevant pinning dynamics manifest only at their surface, 3D geommetry models such as the ones based on the so-called T-A formulation result in the computational modelling of 2D domains~\cite{Wang2019SUST}. These models although represent a significant improvement in terms of the computing time when compared with fully 3D numerical formulations~\cite{Sheng2017IEEE,Tian2021IEEE}, still might suffer of the sufficient level of accuracy for determining subtle changes in the AC-losses of the CORC\textsuperscript\textregistered cables by the change of the twist pitch. Which can in-turn actually modify the vortex orientation inside the SC material, i.e., along the thickness of the SC tape. In other words, 2D-domain simplifications assume that all vortex inside the SC tape are like fluxons with a solid symmetry axis, whilst in the reality these fluxons can be intertwine within the SC material~\cite{Ruiz2009bPRB}. Therefore, despite fully 3D-formulations based on the H-formulations being computationally more expensive, these are certainly needed to reconstruct the full electrodynamics of the CORC\textsuperscript\textregistered cables, in particular when the task ahead is to validate the AC-loss measurements of the early commercial designs of CORC\textsuperscript\textregistered cables~\cite{Majoros2014}. All other models up to date have focused mostly on the validation of their computational principles through comparison of in-house lab-made CORC\textsuperscript\textregistered cables, which has left the door open for investigating further improvements on the full 3D modelling of commercially made CORC\textsuperscript\textregistered cables, to make them affordable with relatively inexpensive workstations.

Therefore, in this paper we present the AC-losses study of the single layer commercial CORC\textsuperscript\textregistered cable~\cite{Majoros2014}, assessing whether their choice of the twist pitch length provides the minimum amount of hysteresis losses. The computational method is based on the H-formulation, enabling a full 3D modelling of the electromagnetic variables and helicoidal geometry of the cable, for which the model itself is validated first with the experimental measurements and analytical solution for a straight 2G-HTS tape in Sec.~\ref{Sec.2} of this manuscript. Then, the CORC\textsuperscript\textregistered cable design is explained with the different twist pitch lengths, where we reveal a direct comparison between our numerical model and the AC-losses measurements reported in the literature. Our calculations show how the optimum twist pitch length can be estimated, by considering different factors such as the intensity of the AC-losses and the amount of 2G-HTS tape involved. In this sense, we disclose the window of change for the twist pitch angle which could offer similar AC-losses at different intensities of an applied magnetic field, confirming the field-ranges for which the pitch angle chosen for the manufacturing of the CORC\textsuperscript\textregistered cables reported in the literature are certainly the most optimal.

\begin{figure}[t]
\centering
\resizebox{0.68\textwidth}{!}{\includegraphics{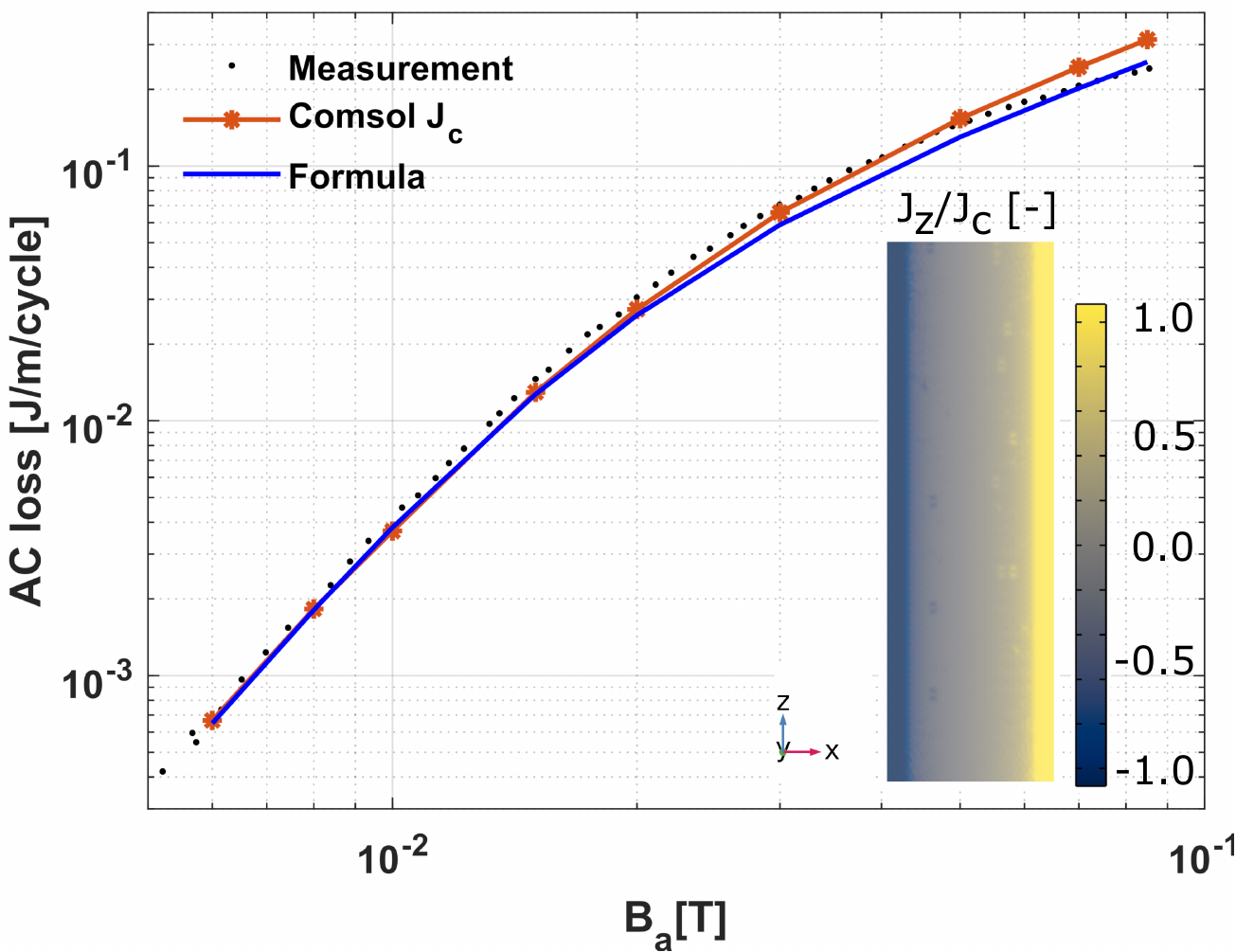}}
\caption{The AC losses comparison of the 12 mm straight tape
measurements, the COMSOL model and analytical solution shown
perfect agreement and confirmed correctness of the model.The inset shown current distribution of the screening current at the peak of the 20 mT magnetic field.}
\label{Fig_1}
\end{figure}


\section{Computational model and validation}
\label{Sec.2}

In this section, we explain and validate our computational model for analysing the AC-losses produced by superconducting wires and cables with a 3-dimensional topology. The model is based on the general PDE formulation of Comsol Multiphysics, which allows a robust implementation of Maxwell equations through 3D Finite Element Methods (FEM), with the state variables defined in what is commonly known as the H-formulation. This formulation allows us to adopt the magneto quasi-steady  (MQS) approach of the critical state theory~\cite{Ruiz2009bPRB} and neglect the time dependence of the electric displacement vector $\textbf{D}$ at the general definition of Amp\`{e}re's law, $\nabla\times\textbf{H}=\partial_{t}\textbf{D}+\textbf{J}$. Assuming the magnetic permeability of any superconductor to be isotropic and equal to the magnetic permeability of free space, $\mu _{0}\approx  4\pi\times 10^{-7}$~H/m. Due to no intrinsic magnetization vector $\textbf{M}$, the magnetic flux density can be simply written as $\textbf{B}=\mu_{0}\textbf{H}$, with $\textbf{H}$ the magnetic field strength vector, and Faraday's law defined by $\nabla\times\textbf{E}=-\mu_{0}\partial_{t}\textbf{H}$. Therefore, if no other material is introduced where the condition $\textbf{B}=\mu_{0}\textbf{H}$ could not be ensured, Faraday's law can be written for the entire space as the vectorial PDE:
\begin{eqnarray}
\label{Eq_1}
\left| \begin{array}{c}
\partial_y E_z - \partial_z E_y \\
\partial_z E_x - \partial_x E_z \\
\partial_x E_y - \partial_y E_x
\end{array} \right| = -\mu_0\left| \begin{array}{c} \partial_t H_x \\ \partial_t H_y \\ \partial_t H_z \end{array} \right|
\, ,
\end{eqnarray}%
where $E_{x}$, $E_{y}$, and $E_{z}$ are the Cartesian components of the electric field $\bf{E}$, and $H_{x}$, $H_{y}$, and $H_{z}$ are the corresponding components of the magnetic field $\bf{H}$. Analogously, we can express Amp\`{e}re's law in the same vectorial notation as follows: 
\begin{eqnarray}
\label{Eq_2}
\left|\begin{array}{c} J_x \\ J_y \\ J_z \end{array} \right|=
\left| \begin{array}{c}
\partial_y H_z - \partial_z H_y \\
\partial_z H_x - \partial_x H_z \\
\partial_x H_y - \partial_y H_x
\end{array} \right| 
\, , 
\end{eqnarray}
where $J_{x}$, $J_{y}$, and $J_{z}$ define the cartesian components of the current density vector $\textbf{J}$. 

Whilst the conductive properties of any non-superconducting material can be introduced by the linear function of Ohm's law, $\textbf{E}=\sigma\textbf{J}$, with $\sigma$ the electric conductivity of the media, in the case of superconductors the material law governing the electrical conductivity of these materials follows a highly non-linear relation. For the sake of simplicity and numerical convergence, this is introduced as an $E-J$ power law. As PDE formulations do not allow instantaneous transitions of state, i.e., current states where any variation of the magnetic field leads to an instantaneous transition of $J=0$ to $J=\pm J_{c}$ when $E\neq0$. Hence, to tackle the non-linearity of Bean's concept~\cite{Bean1962}, and to incorporate the physical principles of the critical state theory~\cite{Ruiz2009bPRB}, a standard $E-J$ power law can be introduced, 
\begin{eqnarray}
\label{Eq_3}
E=E_c \frac{\textbf{J}}{|\textbf{J}|} \left(\frac{|\bf{J}|}{J_c}\right)^n
\, ,
\end{eqnarray}
with $E_{c}$ being the standard critical electric field criterion of $1\mu$V/cm, from which the parameters $J_{c}$ and $n$ can be experimentally measured~\cite{Ruiz2017SUST}. For commercial REBCO tapes at 77 K, $n$ is commonly assumed to be greater than 25 with critical current densities $(J_{c})$ around 25~GA/m$^2$. $J_{c}$ and $n$ can also be as isotropic or anisotropic parameters depending on the requirements of the experimental observations, as it is known that these parameters can depend on external quantities such as magnetic field and temperature. However, it is already known that unless the operational temperature is demanded to change, like in the case of superconducting fault current limiters~\cite{Ruiz2015aIEEE}, isotropic approaches commonly come up with sufficiently accurate results for high transport currents~\cite{Ruiz2019MDPI}. For this reason, within this paper we will demonstrate that the isotropic approach is sufficient for the optimization of CORC\textsuperscript\textregistered cables at different twist pitch angles, and due to it's importance within the cryogenic industry, an accurate estimation of the AC-losses.

As an initial validation of our model, we have confirmed our numerical results through direct comparison with the experimental measurements of the AC losses in a straight YBCO tape under applied magnetic field~\cite{Majoros2014}. The dimensions of the measured tape were 12~mm width, 25~cm length, and a 1~$\mu$m thick superconducting layer with a critical current density of 2.5~GA/m$^2$ at self-filed conditions and 77~K, i.e., with an $I_{c}=300$~A. Then, for tackling the computational challenges associated with the large aspect ratio of the superconducting cross-section, the thickness of this layer and consequently its $J_{c}$ has been renormalized by a factor of 50. This allows to reduce the meshing density around the superconducting tape, increasing  the numerical convergence of the PDE solver whilst reducing its computing time. 

Likewise, our numerical results have been compared not only with the experimental measurements (see Fig.~\ref{Fig_1}), but also confirmed via the analytical solution of Halse~\cite{Halse1970}, 
\begin{eqnarray}
\label{Eq_4}
W=\frac{8\mu_0J_c^2w^2}{\pi}\left[\ln\left(\cosh\left(\frac{\pi H_p}{J_c}\right)\right)-\frac{\pi H_p}{2J_c}\tanh\left(\frac{\pi H_p}{J_c}\right)\right]
\, ,
\end{eqnarray}
where $w$ it the width of the tape and $H_{p}$ is the peak value of the applied (AC) magnetic field. The calculated AC losses by Halse's analytical formula confirms the correctness of our numerical model, allowing us to use it confidently in the modelling of more sophisticated 3D structures such as the CORC\textsuperscript\textregistered cables, whilst gaining an insight of the local distribution of current density along the superconducting tape, as it is shown in the inset of Fig.~\ref{Fig_1}.


\section{Results $\&$ discussion}
\label{Sec.3}

\begin{figure}[t]
\centering
\resizebox{0.68\textwidth}{!}{\includegraphics{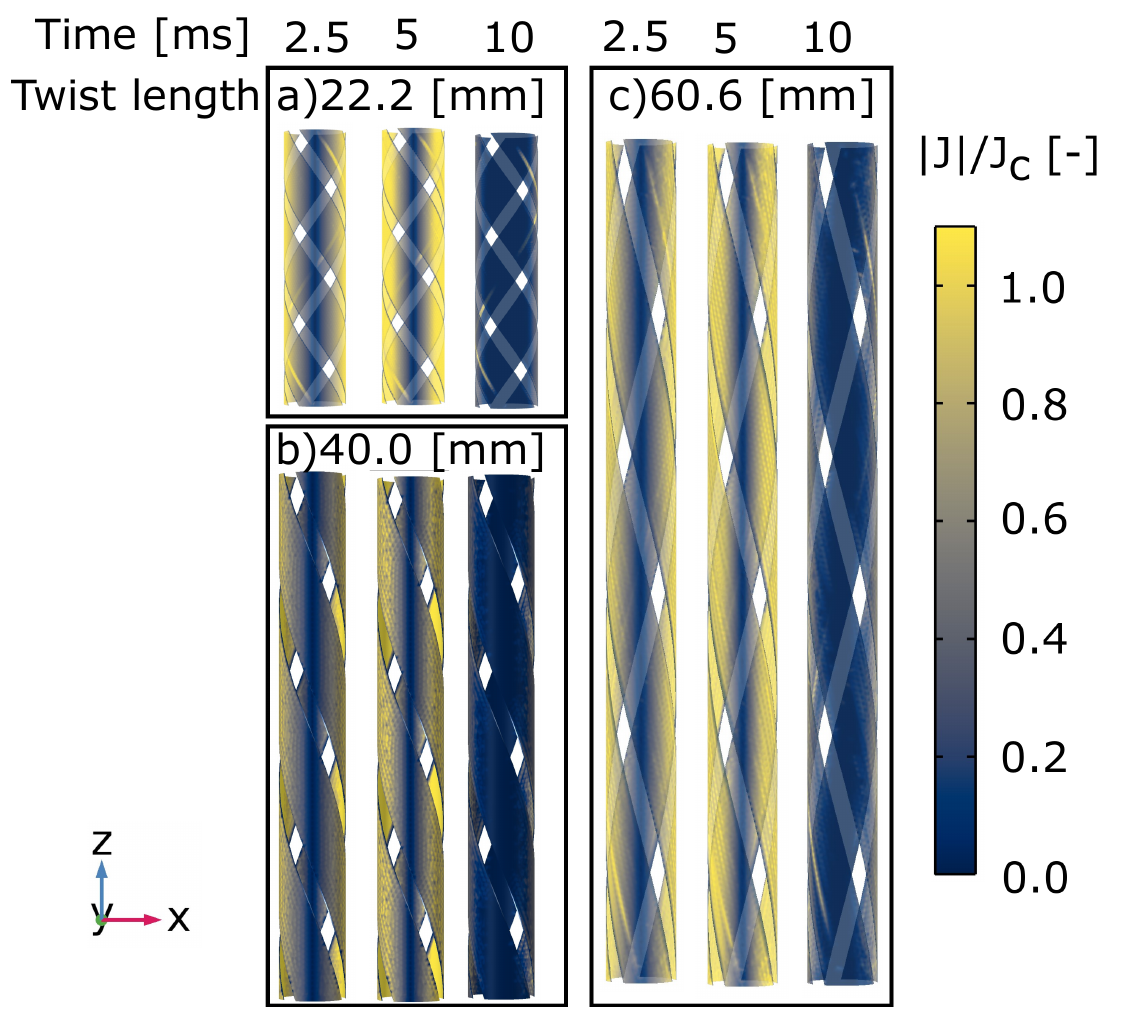}}
\caption{Magnetization currents in CORC cables with twist lengths of (a) 22.2 mm, (b) 40 mm, and (c) 60.6 mm, at half-peak (2.5 ms), first peak (5 ms), and zero-field (10 ms) of an external magnetic field applied in the y-direction $B_{ext}=B_{a}\sin(\omega t)$.}
\label{Fig_2}
\end{figure}

In this section, we focused at first on the modelling of the AC-losses of experimentally measured 1-layer CORC\textsuperscript\textregistered cable~\cite{Majoros2014},  validating the effectiveness of our model. Then, a comprehensive study on how the twist pitch length influences the calculated AC-losses has been pursued, concluding whether the manufactured cable design can be considered as optimal or not.

\begin{figure}[t]
\centering
\resizebox{0.68\textwidth}{!}{\includegraphics{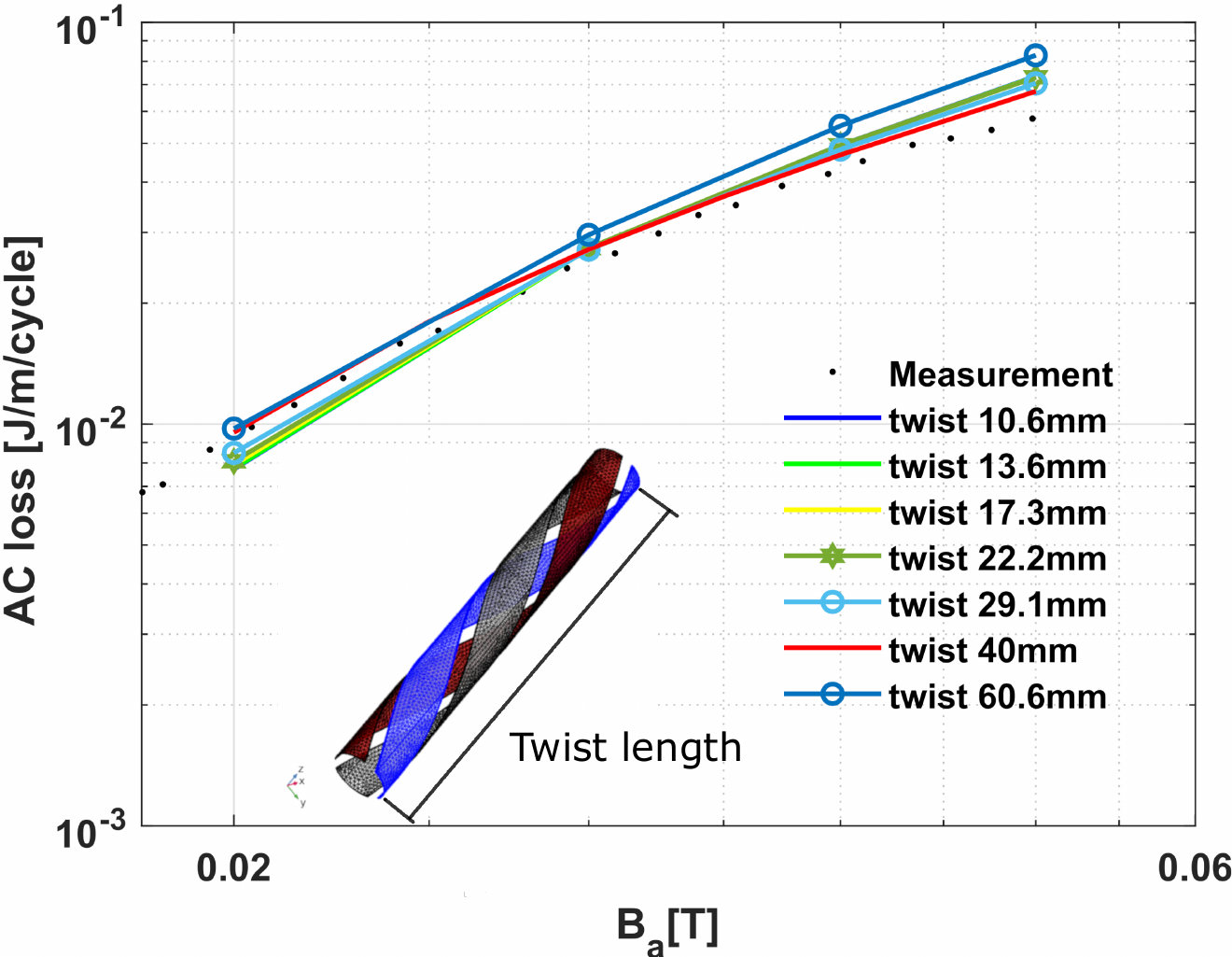}}
\caption{The AC losses of the CORC\textsuperscript\textregistered cable with 40 mm twist pitch
length shown nice agreement with measurement. The study of different twist pitch confirmed the optimum twist length of 40 mm or twist angle $68.72\degree$ with slight increase and reduction of the AC losses by 15$\%$ outside of this length. The inset shows the twist pitch length in the cable.}
\label{Fig_3}
\end{figure}

The full geometry of a 3D 1-layer CORC\textsuperscript\textregistered cable with different twist pitch lengths has been modelled with 3 tapes twisted along the $z$ axes on an stainless-steel former of 4.96 mm diameter, shown in Fig. \ref{Fig_2}. Since the experimental measurements have shown negligible eddy current in the stainless-steel former~\cite{Majoros2014}, this can be assumed as an electrically insulated material in the computational model, with the same properties as the surroundings of the superconducting tapes. The tapes in the cable are 4 mm wide with a critical current of $I_{c}=25$~A per mm width, measured at $77$~K and self-field conditions. In the model we applied the oscillating magnetic field in the $y$ direction, with amplitudes ranging from $20$~mT to $50$~mT based on the original experimental conditions. The tapes are initially arranged with a twist pitch angle of 68$\degree$,  i.e., for building a CORC\textsuperscript\textregistered cable of $40$~mm along the $z-$length. Our simulations show an excellent agreement with the experimental measurements (Fig.~\ref{Fig_3}), with slightly higher losses at fields greater than $40$~mT. It has been reported in ~\cite{Majoros2014} that their experimental measurements beyond this field magnitude might have been affected by some heating effects, which may indicate a decrease in the $J_{c}$ and consequently the AC-Losses. For this reason, at the field range when full penetration of the magnetic field occurs, our numerical simulations can be considered reliable for not only the the AC-losses, but also because it confirms the negligible influence of eddy currents at the stainless steel former. It is also worth checking the distribution of magnetization currents due to their influence on the occurrence of power losses. The current distribution of this case is shown in the CORC\textsuperscript\textregistered cable at the left bottom pane of Fig.~$\ref{Fig_2}$, where the screening currents spread into the tapes in a similar way to what can be seen in a straight tape (see Fig.~\ref{Fig_1}) or a bulk wire~\cite{Ruiz2012APL,Ruiz2013JAP}, i.e, from the outer edges of the cable (perpendicular to the direction of the applied magnetic field) towards its center. Notice that in Fig.~\ref{Fig_2}, it is the normalized magnitude of the total current density $|\bf{J}|$ what is shown. Although the dominant component of the current density is $J_{z}$, the helix structure of the CORC\textsuperscript\textregistered cable does not allow a greater comparison with the above mentioned `2D' geometries as the one shown here. 

Having verified our numerical model for the 1-layer 3 tapes CORC\textsuperscript\textregistered cable, its twist pitch angle  has been systematically changed from $\sim35\degree$ to $\sim75\degree$. This has helped us to determine whether the twist pitch angle of the experimentally measured cable with twist length $\Gamma_{l}=40$~mm, and diameter $D=4.96$~mm, i.e, $\beta=\arctan{\left(\Gamma_{l}/(\pi D)\right)}=68.72\degree$, can be considered as an optimal CORC\textsuperscript\textregistered cable design or not. Therefore, the results presented in Fig.~\ref{Fig_3} are for showing the AC-losses of different CORC\textsuperscript\textregistered cables considering a maximum rate of change of $10\%$ the pitch angle reported in the experiments.  This is equivalent to a study of at least 7 twist lengths, with pitch angles of approximately $34.36\degree$, $ 41.23\degree$, $48.10\degree$, $54.97\degree$, $61.85\degree$, $68.72\degree$, and $75.59\degree$ respectively. Likewise, the spiral length, i.e, the approximate amount of coated conductor used for a given twist pitch length can be calculated by $CC_{l}=\sqrt{\Gamma_{l}^{2}+(\pi D)^{2}}$, which results in significant differences in the amount of of tape required for each of the twist pitch angles analyzed, or being more specific, three times $18.88$~mm,  $20.72$~mm, $23.34$~mm,  $27.15$~mm, $33.01$~mm, $42.93$~mm, and $62.6$~mm, respectively.

Based on this study we can conclude that for the broad range of twist pitch lengths  studied, the distribution and dynamics of magnetization currents in the 1-layer 3-tapes CORC\textsuperscript\textregistered cable manufactured by Advanced Conductor Technologies (see Fig.~\ref{Fig_2}), and consequently its AC-Losses (see Fig.~\ref{Fig_3}), do not significantly change as a function of the twist pitch angle. However, by shortening the twist pitch angle, i.e., by reducing the cable-length, also called the twist pitch length, the AC-losses at high field (50~mT) increase in about a $15\%$. Whilst at low field (20~mT) they tend to decrease in the same ratio, at least for $\Gamma_{L}<40$~mm. Despite a reduction of the AC-losses at low field being beneficial, the reduction in the distance covered by the CORC\textsuperscript\textregistered cable is of at least $27\%$ to $73\%$ the length covered by the manufactured cable with $\Gamma_{l}=40$~mm. Therefore, in order to cover the same $40$~mm distance, the amount of tape required by the CORC\textsuperscript\textregistered cable with $\beta\simeq 34.36\degree$ and $\Gamma_{l}\simeq 10.65$~mm, would be of at least 2.27 times the amount of tape required for the standarized $40$~mm cable with $\beta=68.72\degree$. 

The slight reduction in the the AC-losses at low magnetic fields and low pitch angles is due to the reduced path in the loops of current density that form along the $z-$axis, i.e, screening the magnetic field applied along the $y-$direction. However, as the applied magnetic field increases, for shorter loops of the magnetization currents, i.e., for lower twist pitch angles, larger becomes the magnetic field penetration rendering  to higher AC-losses. Therefore, a certain balance must be found between the twist pitch length and the AC-losses of the CORC\textsuperscript\textregistered conductor, such that for long cable lengths and high magnetic fields, no  excessive amount of 2G-HTS is demanded. As the AC-losses of the CORC\textsuperscript\textregistered cable for $\Gamma_{L}>40$~mm and applied magnetic fields greater than $40$~mT start to show a noticeable increment (see Fig.~\ref{Fig_3}), we have found that the optimal twist pitch length for the 1-layer 3-tapes CORC\textsuperscript\textregistered cable manufactured by Advanced Conductor Technologies is indeed $40\pm5$~mm. In fact, as the amount of conductor required for $\Gamma_{l}=40$~mm is of just $42.93$~mm, no more than a $3\%$ increment in the cabling cost will be expected (excluding the cost of the SC-former) in comparison with an equivalent cable composed by straight tapes, making this dimensions ideal for the inception marketing of CORC\textsuperscript\textregistered cables in high power systems applications. 

\section{Conclusion}~
\label{Sec.4}

Based on the possibility to validate the H-formulation in a fully 3D model of straight SuperPower 4-mm width tapes (SCS4050) either from experimental observations or by fully analytic methods, we expanded our model to present a comprehensive study on the twist pitch angle of CORC\textsuperscript\textregistered cables. Here we shown a CORC\textsuperscript\textregistered cable model with 3 tapes in 1-layer, following the exact manufacturing dimensions and physical properties of the cables assembled by Advanced Conductor Technologies LLC. We have proven how our method reproduces with good accuracy the experimental measurements of AC-losses when the cable is exposed to a transverse magnetic field, it without the need for reducing the dimensionality of the superconducting tapes from a finite thickness layer to a strip approach. Seven CORC\textsuperscript\textregistered cables with different twist pitch lengths have been studied in order to confirm their optimum design.
Twist pitch lengths in the range of $\Gamma_{L}=10.6-60.6$~mm show a slight increase of the AC-losses by a maximum of 15$\%$ at high magnetic fields, and a similar reduction by 15$\%$ at low magnetic fields. This reduction of the AC-losses at low fields and $\Gamma_{L}<40$~mm is caused by shorter screening currents loops. With an increasing magnetic field, the AC-losses escalated above the measured AC-losses value in the 40 mm cable case. This was due to higher amount of the HTS tapes per the cable length. In the case of the twist length 60.6 mm, the AC-losses increases due to the current loops being too long. This study has confirmed the optimum twist pitch length is indeed the one in the experimentally measured cable, i.e., of 40 mm. In fact, as by shortening the twist pitch length the tape length required increases, then we have found that an optimal twist pitch length of $40\pm5$~mm is encountered by balancing the AC-losses at high and low field encountered, with the need to have a sufficiently long twist pitch length for the manufacturing and commercial production of the CORC\textsuperscript\textregistered cable. Further studies on multiple layer CORC\textsuperscript\textregistered cables are required to confirm whether the above conclusions stand as a principle of generality, as although from the electromagnetic point of view the physical model will remain unaltered, the aggregated computational burden might present significant challenges for the uncovering of new physics on these designs.

%
\section{References}\vspace*{5mm}

\bibliographystyle{iopart-num}

\bibliography{References_Ruiz_Group}

\section*{Acknowledgements} This work was supported by the UK Research and Innovation, Engineering and Physical Sciences Research Council (EPSRC), Grant Ref. EP/S025707/1 led by H.S.R. All authors acknowledge the use of the High Performance Computing facility ALICE at the University of Leicester.

\end{document}